\documentclass[namedreferences]{solarphysics}

\usepackage[optionalrh,solaromanenum]{spr-sola-addons} 
\usepackage{natbib}
\usepackage{graphicx}                    
\usepackage{amssymb}                    
\usepackage{color}                       
\usepackage{url}                         
\usepackage{epstopdf}
\usepackage{color}
\usepackage{enumerate}
\usepackage{bm}

\usepackage{solaheader}
\newcommand{\solareceiveddate}{22 February 2015} 
\newcommand{\solaaccepteddate}{11 July 2015}
\renewcommand{\solayear}{2015}
\newcommand{\ld}{\,--\,}

\begin{document}

\begin{article}

\begin{opening}

\title{Enhanced Acoustic Emission in Relation to the Acoustic Halo Surrounding Active Region 11429}

%

\author[addressref={aff1,aff2},email={email:hanson@mps.mpg.de}]{\inits{C.S.}\fnm{Chris S.}~\lnm{Hanson}}
\author[addressref={aff1},corref]{\inits{A.C.}\fnm{Alina C.}~\lnm{Donea}}
\author[addressref={aff3},corref]{\inits{K.D.}\fnm{K.D.}~\lnm{Leka}}
\address[id=aff1]{School of Mathematical Sciences and Monash Centre for Astrophysics, Monash University, Melbourne, Victoria 3800, Australia}
\address[id=aff2]{Max-Planck-Institut f\"ur Sonnensystemforschung, Justus-von-Liebig-Weg 3, 37077 G\"ottingen, Germany}
\address[id=aff3]{NorthWest Research Associates, 3380 Mitchell Ln. Boulder, CO 80301, USA}

\runningauthor{C.S. Hanson, \textit{et al.}}

\begin{abstract}
The use of acoustic holography in the high-frequency $p$-mode spectrum can resolve the source distributions of enhanced acoustic emissions within halo structures surrounding active regions. In doing so, statistical methods can then be applied to ascertain relationships with the magnetic field. This is the focus of this study. The mechanism responsible for the detected enhancement of acoustic sources around solar active regions has not yet been explained.  Furthermore the relationship between the magnetic field and enhanced acoustic emission has not yet been comprehensively examined. We  have used  vector magnetograms from the \textit{Helioseismic and Magnetic Imager} (HMI) onboard the \textit{Solar Dynamics Observatory} (SDO) to image the magnetic-field properties in the halo. We have studied the acoustic morphology of an active region, with a complex halo and ``glories'', and we have  linked some  acoustic properties to  the magnetic-field configuration. In particular, we find that acoustic sources are significantly enhanced in regions of intermediate field strength with inclinations no different from the distributions found in the quiet Sun. Additionally we have identified a transition region between the active region and the halo, in which the acoustic source power is hindered by inclined fields of intermediate field strength. Finally, we have compared the results of acoustic emission maps, calculated from holography, and the commonly used local acoustic maps, finding that the two types of maps have similar properties with respect to the magnetic field but lack spatial correlation when examining the highest-powered regions.

\end{abstract}

%

\end{opening}
\section{Introduction}
\label{S-Introduction} 
The enhanced acoustic power surrounding active regions has been the focus of many recent observational and numerical studies. These enhanced regions are commonly referred to as {\it acoustic halos} and typically have a power that exceeds that of the quiet Sun by $40$\ld$60\%$ \cite[\textit{etc.}]{hindman_brown_1998}.  These halos were first observed near the photosphere by \citet{brown_etal_1992}, and have since been observed at chromospheric heights \citep{braun_etal_1992,toner_labonte_1993}. Thus far, the mechanisms responsible for this phenomenon are not well understood. With regards to the morphology, \citet{khomenko_collados_2009} provide a concise summary, and the key points are:
\begin{enumerate}[i)]
\item a disproportionate enhancement of acoustic power occurs in the $5.5$\,--$7.5$\,mHz frequency range 
\item the halos occur in low to intermediate fields (50\ld300\,G) \citep{hindman_brown_1998}, and diminish rapidly with greater field strengths
\item enhancement is aided by regions of near-horizontal field inclination, in particular between locations of opposite polarity \citep{schunker_braun_2011,rajaguru_etal_2012}

\end{enumerate}

Numerous theories have suggested  possible mechanisms responsible for the halo phenomenon. In simulations, \citet{jacoutot_etal_2008} showed that high-frequency turbulent convective motions, in the presence of moderate magnetic fields, may enhance the local acoustic emission. Another study by \citet{khomenko_collados_2009} proposed that the refraction of  fast-waves in the higher atmosphere could deposit additional energy into photospheric regions of intermediate field strength. Also, \citet{kuridze_etal_2008} suggested that high-order azimuthal modes may become trapped under the magnetic canopy, enhancing the observable vertical-velocity power. Meanwhile \citet{hanasoge_2008,hanasoge_2009b} argued that enhanced acoustic oscillations may be the result of a MHD mode-mixing (scattering).  While these theories all propose different mechanisms, further observational statistics are required in order to test and examine their validity. 

Typically in observational studies, the acoustic power is determined from the local Doppler signature at a given point on the solar surface. It is then a standard procedure to use these Doppler images to generate and interpret power maps at frequencies between 2\ld10\,mHz. In these local Doppler power maps, the acoustic waves ($p$-modes) generated by convective motions, as well as any additional acoustic source mechanisms, will be seen. In order to differentiate the additional acoustic sources, apart from the background oscillations, the computational acoustic holography  technique is applied \citep{lindsey_braun_1997}. Power maps generated in holography show the acoustic sources within the halo that generate outgoing waves, which reemerge some distance away. 

Helioseismic holography \citep{lindsey_braun_2000_review} was fashioned expressly as a local discriminator of helioseismic anomalies, whether of sunspots and plages \citep{braun_lindsey_2000b,braun_lindsey_2000a}, seismic transients from flares from the near solar hemisphere \citep{lindsey_donea_2008}, active regions on the Sun's far hemisphere \citep{lindsey_braun_2000,braun_lindsey_2001}, or magnetic anomalies submerged in the solar interior \citep{lindsey_braun_1997,braun_etal_1992}. The basic principles and applications of helioseismic holography are described by \citet{lindsey_braun_2000_review}. The technique conceptually entails the extension of the basic principles of electromagnetic optics to treat signatures of acoustic radiation on the Sun's surface, in a manner similar to how our eyes treat the electromagnetic radiation at the surface of the cornea.

 Typically, enhanced acoustic-source halos occur around large complex multi-polar regions \citep{donea_newington_2011}, but have also been used  to search for additional emission of acoustic waves relative to the quiet Sun \citep{donea_2000}. \citet{donea_2000}  also found that acoustic emission, both in the magnetic peripheries and in the quiet Sun, appears to maintain a temporal character no more episodic than random Gaussian noise. Within these emission halos, discrete regions of sustained intense acoustic power can also be observed, and these are known as {\it acoustic glories} \citep{braun_lindsey_1999,donea_2000}. 
 
Utilizing the data available from the \textit{Solar Dynamics Observatory} (SDO) mission, in particular the onboard \textit{Helioseismic and Magnetic Imager} (HMI), we present detailed findings of the active region AR\,11429, which possesses an encompassing enhanced acoustic-emission halo. This acoustic-source halo is examined through the computation of egression (outgoing) power maps and is compared to the commonly examined local acoustic-power halo. While recent studies by \citet{schunker_braun_2011} and \citet{rajaguru_etal_2012} have examined the relationship between local acoustic power and the magnetic field, none has yet investigated the relationship with the acoustic-source enhancement of the halos and glories. Additionally, the similarities between these two types of halos have not yet been extensively examined. In this article we will  examine both the local oscillation power (referred herein as {\it local acoustic power}) and the acoustic power that is emitted from a point and resurfaces some distance away (refereed herein as {\it acoustic source power}). We compare these with various parameters, including the magnetic-field morphology,  in order to further understand the acoustic morphology of complex active regions.

The active region AR\,11429 has been chosen for an investigation because of its magnetic-field complexity. The active region generated numerous flares, including two X-class flares, and it is interesting in that its polarities are reversed \citep{elmhamdi_etal_2013}. Specific to this study, this region is interesting because of the high levels of acoustic radiation from flares and surrounding magnetic features \citep{donea_hanson_2013}. However, we note that previous studies have examined non-flaring regions and found that high levels of acoustic source power surround the magnetic regions \citep[\textit{e.g.}][]{donea_newington_2011}. In this article we expand upon these results, with particular focus on the relationship between the geometry of the magnetic field, acoustic-emission halos and glories.

The article is organised in the following fashion: Section~\ref{S-Obs} outlines the analysis performed on the data as well as the statistical precautions we have taken with the magnetic data. Section~\ref{S-power} presents the computed power maps of the active region with an overview of the halo morphology. Section~\ref{S-results} will state the results of the analysis and comparison of the two halos, with Section~\ref{S-Discussion} comparing the results with similar studies, and concluding on the findings of this article.


\section{Observations}
\label{S-Obs} 

We have analysed the NOAA Active Region (AR) 11429 over 24 hours on 8 March 2012,  during which the active region passed through the central meridian at 22:00 UT. All data products are from the \textit{Helioseismic and Magnetic Imager} \citep[HMI:][]{scherrer_etal_2012} onboard the \textit{Solar Dynamics Observatory} (SDO). Specifically, we have performed analysis on the Doppler velocity [$v$] and disambiguated magnetic-field components ($\bm{|B |}$, inclination, azimuth).  The Doppler measurements have a cadence of 45 seconds, while the field components (SHARPs products, \textsf{hmi.sharps$\_$720s.1449}) have a cadence of 720 seconds. For this study all products are remapped using Postel projection routines, with the centre of the active region at the centre of the maps. The resultant tracked Doppler cube is then Fourier transformed in time and filtered into 1\,mHz bandwidths.

We note that in the quiet Sun there are spatial fluctuations, caused by the interference among modes with the same frequency, that can appear in local acoustic power maps. \citet{chou_etal_2009} stated that the only way to reduce the errors that arise from these fluctuations is to increase the number of observed frames. Therefore, 24 hours of observations ensures that a significant number of frames are taken into account.

\subsection{SDO Vector Magnetograms}
\begin{figure}
\centering
\includegraphics[width=\textwidth]{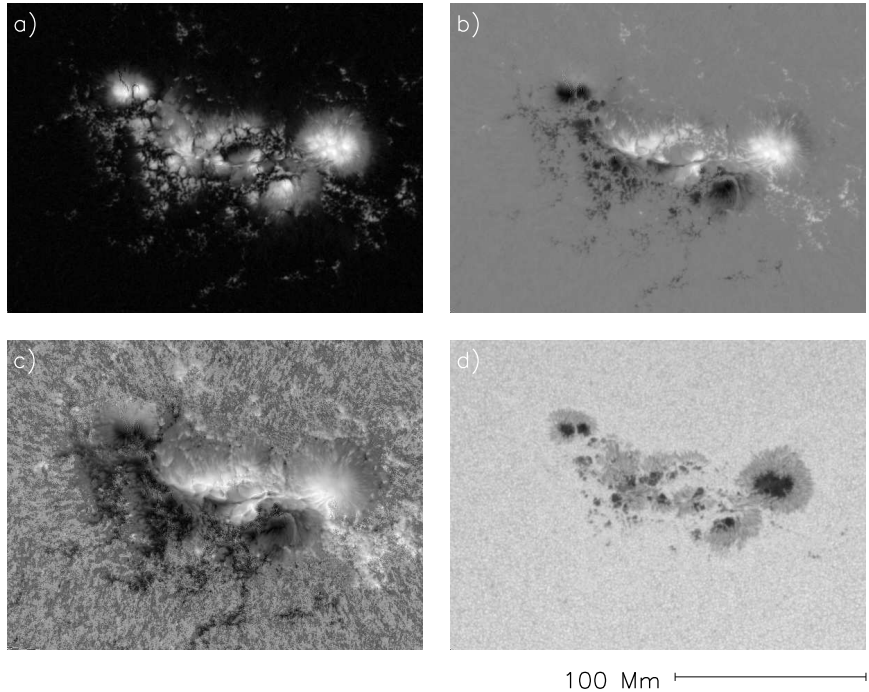} 
\caption{8 March 2012 AR\,11429. The magnetic-field components are: a) the $|\bm{B}|$ (scaled from 0 to 2500 G), b) $B_{\textrm{los}}$ (scaled between $\pm2500$~G) and c) Field Inclination $\gamma$ ($\pm 90^{\circ}$). The intensity continuum image of AR\,11429 is shown in panel d).}
\label{fig:magneticmaps}
\end{figure}

The SHARPs products are computed through the inversion of the Stokes parameters ($I, Q, U, V$) within the SDO pipeline, which utilizes a custom version of the {\it Very Fast Inversion of the Stokes Vector} \citep{borrero_etal_2011,centeno_etal_2014}. The inherent $180^\circ$ ambiguity of the azimuth is also resolved in the HMI pipeline, using a minimum-energy method \citep{metcalf_1994,leka_etal_2009,hoeksema_etal_2014}.   Subsequently, we derive the magnetic-field components ($B_x,B_y,B_z$) from the HMI line-of-sight observations of the field $\bm{|B|}$, inclination and disambiguated azimuth. These components are relative to the image plane (line-of-sight), and as such we have performed the appropriate transforms to describe the field vectors in relation to the heliographic plane (see \citet{gary_hagyard_1990}). We note that although the active region passes through the central meridian (hence $B_x$ requires little transform at that time), it is still North of disk centre and thus requires more significant transforms to $B_y$ and $B_z$. Having calculated the heliographic-field components, we then define the inclination [$\gamma$] in the same manner as  \citet{schunker_braun_2011},

\begin{equation}
\textrm{tan}(\gamma)=\frac{B_z}{B_h},
\end{equation}
where $B_h=\sqrt{B_x^2+B_y^2}$. Here $\gamma$ ranges from $-90^{\circ}$ to $+90^{\circ}$, with the horizontal field being $0^{\circ}$ and $+\gamma$ outward facing from the surface. Figure\,\ref{fig:magneticmaps} shows the vector magnetograms and continuum maps for AR\,11429. 

The accuracy of the inferred field vector depends upon the strength of the $I,Q,U,V$ polarization signals. In the quiet Sun, where the acoustic halo is present, the signal strength of the Stokes parameters is generally weak. This results in a higher noise level.
Since we have been primarily concerned with the surrounding quiet Sun, we have taken precautions when interpreting the field information within the halo. We first reduced the noise in the azimuth data by taking an average of the field vectors over a period of 36 minutes. Longer averaging results in additional uncertainty due to solar evolution. In the results of this article we have used the magnetic field when the active region is on the central meridian. For consistency, we have examined the magnetic field at different times, and have found the results to be in agreement. The second precaution was to consider only pixels that have a good signal to noise ratio ($|\bm{B}|^2/|\bm{B}_\textrm{error}|^2>1$). Finally, we have analysed a quiet Sun patch (void of any apparent magnetic structure) finding that a minimum field strength of 60\,G remains. As such we omit pixels with field strength below 60\,G, attributing these values to noise. Upon performing these precautions, the remaining pixels number approximately $50\,\%$ of the original amount within the halo. However, as a result the confidence in interpreting these pixel values is high. 


\section{Calculating Cospatial Local Acoustic and Acoustic Emission Power Maps}\label{S-power}

This study compares computed local acoustic power maps and acoustic-source power maps. We compute local acoustic power maps from the HMI observations of the Doppler velocity [$|\psi (\bm{r},t)|^2$], while the acoustic-source power maps are derived using helioseismic holography \citep{lindsey_braun_1997}. Holography uses the convolution of a Green's function [$G_{+}$] to time reverse the observed acoustic wave field $\psi(\bm{r}',t')$ to a location and time $(\bm{r},z,t)$ at depth $z$. This in turn, renders a regressed acoustic wave field (termed the {\it Egression} in previous literature):
\begin{equation}
H_+(z,\mathbf{r},t)=\int \textrm{d}t' \int\limits_{a<|\mathbf{r}-\mathbf{r'}|<b} \textrm{d}^2\mathbf{r}G_{+}(z,\mathbf{r}',t',\mathbf{r},t)\psi(\mathbf{r}',t).
\end{equation}
In this study, calculations are made over an annular pupil of inner radius 7~Mm and an outer radius of 45~Mm, at the surface ($z=0$). The acoustic-source maps have a spatial resolution of 2.5 Mm, a diffraction limit attained by applying a diagnostic with an annular pupil. For comparative purposes, both the power maps of the local acoustic [$|\psi(\bm{r},t)|^2$] and acoustic-source [$|H_+(0,\bm{r},t)^2]$, referred herein as $|H_+|^2$) power  are normalised to the quiet-Sun values [$|\psi_0|^2$ and $|H_{+0}|^2$, respectively]. 

The panels in Figure~\ref{fig:powermaps} show the integrated acoustic-source power (left column) and the local acoustic power maps (right column) in the frequency bands centred at 3, 6, and 9\,mHz. Both acoustic-source power and local acoustic power are seen to be suppressed in the active region at all frequencies. This is coincident with the location of moderate to strong magnetic fields, and includes plages. At 6~mHz a power excess can be seen in both power maps surrounding the active region.

For statistical purposes we define the acoustic-source halo as the region of enhanced power surrounding the active region at 6\,mHz. Specifically, we choose to define the halo as the region with an acoustic-source power 
 that exceeds $140\,\%$ of the source power in the quiet Sun ($|H_+|^2/|H_{+0}|^2$; see the yellow contour of Figure~\ref{fig:haloandg}). Our choice in this minimum gives a definable boundary for computing masks within which the pixels can be examined. If a lower limit  were chosen the halo becomes more diffuse with the quiet Sun. 
Additionally, the area having intermediate source power ($1\le|$H$_+|^2/|$H$_{+0}|^2<1.4$) seems to be more of a transition area, where the magnetic field (near the active region) is strong. We name this area the {\it seismic plateau}, and we discuss its properties later. 

Small regions within the halos where the excess in acoustic source power is more then $1.8\times ~|H_{0+}|^2$ have been identified as {\it acoustic glories}. These regions are located in the halo and are delimited by the black contours in Figure~\ref{fig:haloandg}. The choice in this lower limit is based on a visual inspection of the source maps, with the identification of the most conspicuous areas. 


\subsection{Halo Morphology}

We are interested in the morphology, and respective micro-structure, of the source and local acoustic halos and whether there is any correlation between the two and other observable values.

\begin{figure}
\centering
\includegraphics[width=\textwidth]{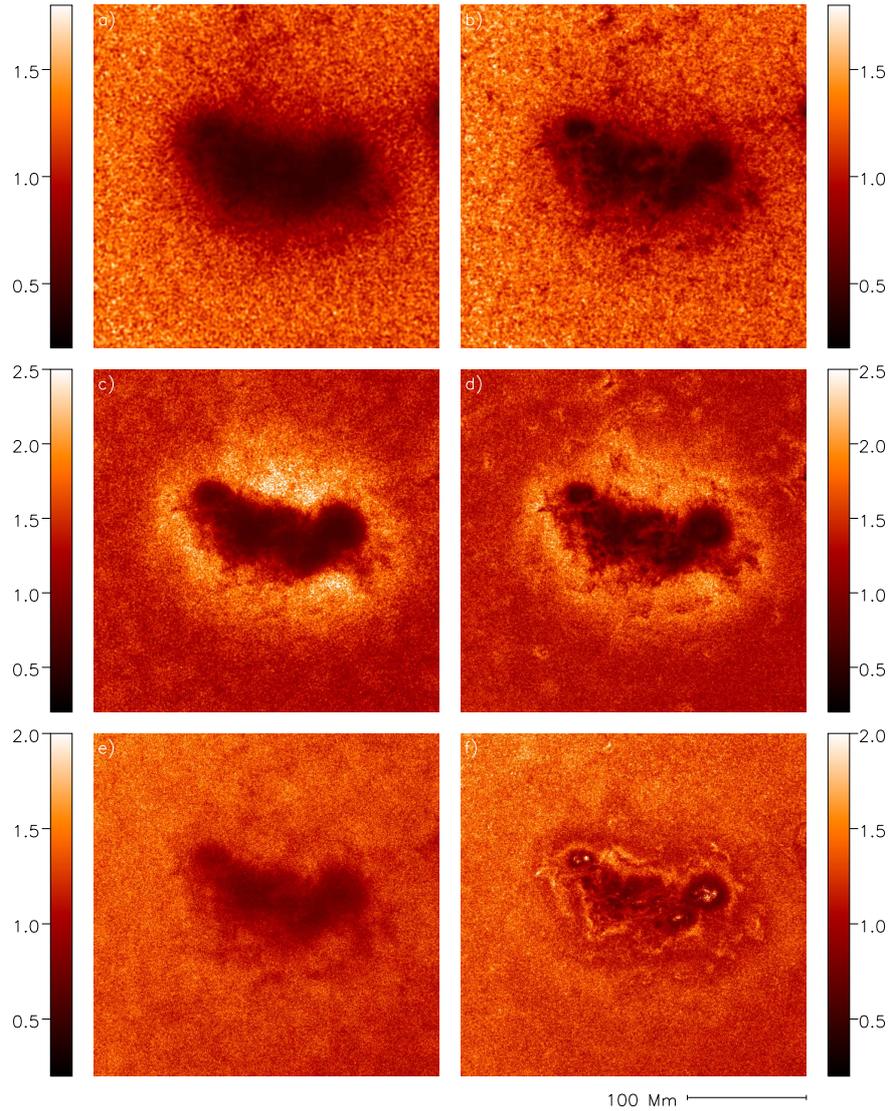} 
\caption{Acoustic-source (left) and local acoustic power (right) of AR\,11429, averaged over 24 hours. From top to bottom the frequencies of the power maps (1\,mHz bandwidth) are centred at 3, 6, and 9\,mHz. The maps are normalised to the quiet Sun. 
}
\label{fig:powermaps}
\end{figure}

The central deficit in the power maps of our active region (Figure~\ref{fig:powermaps}), which correlates with the strong magnetic fields, has a power approximately $0.1$ of quiet Sun values. Between 2.5\ld3.5\,mHz (panels {\it a} and {\it b}) a region of intermediate power deficit ($\approx0.7$\,quiet Sun units) extends from out from this central region ranging in thickness (from penumbra) between 27\ld60\,Mm. This deficit is known as an {\it acoustic moat} \citep{lindsey_braun_1998a}. 

At 6\,mHz the area surrounding the active region that was occupied by the 3\,mHz acoustic moat is now a region of enhanced acoustic-source power($\approx1.6$ quiet-Sun units), which is detected in the helioseismic holography maps. This enhancement is also seen in the local acoustic power maps ($\approx1.5$ quiet-Sun units). The halo surrounds the active region with a width of 25\ld60\,Mm, and in some regions slightly extends beyond the region suppressed at 3\,mHz.

The bottom two panels of Figure~\ref{fig:powermaps} show that at higher frequencies (9\,mHz) the acoustic-source halo has become diffuse with the surrounding quiet Sun, while the region of power deficit  corresponding to the strong magnetic fields remains. However, the local acoustic power halo is now more compact (10\,Mm wide) outlining the active region, with a hint of some diffuse enhancement beyond that ($100$\,Mm). There are also some bright compact acoustic power spots that can be seen within the central parts of the active region. These large power spikes (and their cause) have often been ignored in studies, and have only recently begun to be addressed \citep{zharkov_etal_2013,donea_lindsey_2015}. As the power halos within the surrounding quiet Sun are the focus of this study, we too will not analyse these spikes.

Magnetic plages are located in concentrations around the penumbral regions and are known to be strong absorbers of the local acoustic power \citep{braun_etal_1990}. As a result, both halos are non-uniform in their power distribution, with power suppression in the plages ($\approx0.9$ quiet-Sun units). Close examination shows that local acoustic-power suppression is more sensitive to the magnetic plage than the acoustic-source power. This suggests that the surface amplitude [$\psi(\mathbf{r},t)$] is strongly altered in the presence of a magnetic field.


\subsection{Excess Acoustic-Source Power: Glories}
\begin{figure}
\centering
\includegraphics[scale=1.8]{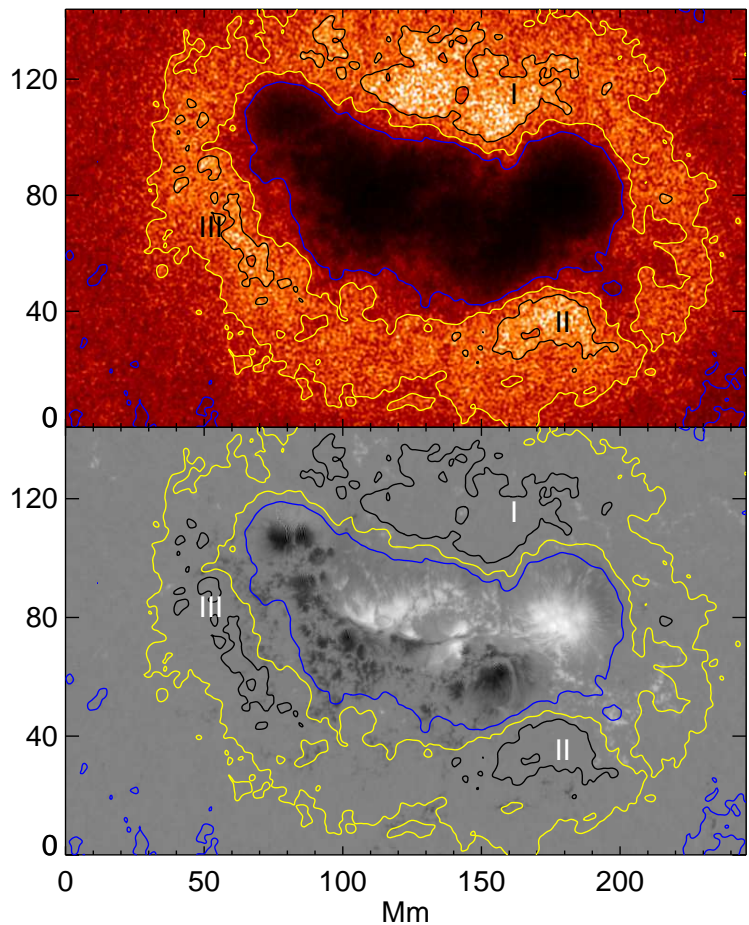}
\caption{Top: Acoustic-source power map (6~mHz) with contours surrounding the glories (black), halo (yellow) and the inner boundary of the plateau (blue). We specify the halo as having 1.4 times more acoustic source power than that of the quiet Sun, with the glories larger than 1.8 times. The inner boundary of the seismic plateau [$|H_+|^2/|H_{+0}|^2=1$] is outlined by the blue contour and extends to the inner boundary of the halo. We number the glories (I, II, III) for clarity in further results. The same contours are overlaid onto maps of $B_z$ (bottom), which indicates the absence of magnetic fields in enhanced acoustic source regions. }
\label{fig:haloandg}
\end{figure}

\begin{table}
\begin{tabular}{llcccccc}
		&Region				&	I 	 & II 	& III 	& Halo 	& Plateau & QS \\\hline
		&3mHz ($\pm 0.25$)		&	0.68&	0.59&	0.73&	0.68&	1.03	&	1\\
Acoustic-Source
		&6mHz	($\pm0.16$)		&	1.98&	1.95&1.91	&	1.65&	1.10	&	1\\
		&9mHz	($\pm0.10$)		&	1.07&	1.07&	1.12&	1.02&	1.11	&	1\\\hline

		&3mHz	($\pm0.13$)		&	0.80&	0.71&	0.87&	0.78&	1.07	&	1\\
Local Acoustic
		&6mHz	($\pm0.20$)		&	1.70&	1.70&	1.74&	1.50&	1.09	&	1\\
		&9mHz	($\pm0.16$)		&	0.99&	0.99&	1.05&	1.03&	1.15	&	1\\\hline
Magnetic Field
		&$|\bm{B}|$ ($\pm27$~G)				&	90	&	94	&	84	&	130	&	267		&	60\\
		&$\gamma\ (\pm10^\circ)$	&	-6	&	-7	&	-13	&	-7	&	-12		&	2\\

\end{tabular}
\caption{The quiet-Sun normalised acoustic power and magnetic properties of six distinct regions of acoustic-source power}
\label{table:summary}
\end{table}
To discriminate regions of different acoustic emissivity, we identify three regions, I, II, and III, in the acoustic-source halo that are significantly enhanced relative to the halo (see black contours in Figure~\ref{fig:haloandg}). Table~\ref{table:summary} shows a summary of the total local acoustic/acoustic-source power averaged over the defined areas, along with estimates of the mean magnetic-field strength and inclination. These regions are clear of magnetic plages  and have high acoustic power. Interestingly, the local acoustic power halo does not show a similar excess power at the glories' location as the source power, although there is some enhancement ($\approx1.7$ quiet-Sun units).  Remarkably there is little difference in the power between each of the three regions, in their respective acoustic-source and local oscillation power. Average inclination angles also show the horizontal nature of the magnetic field within these regions.

\begin{figure}
\centering
\includegraphics[width=\textwidth]{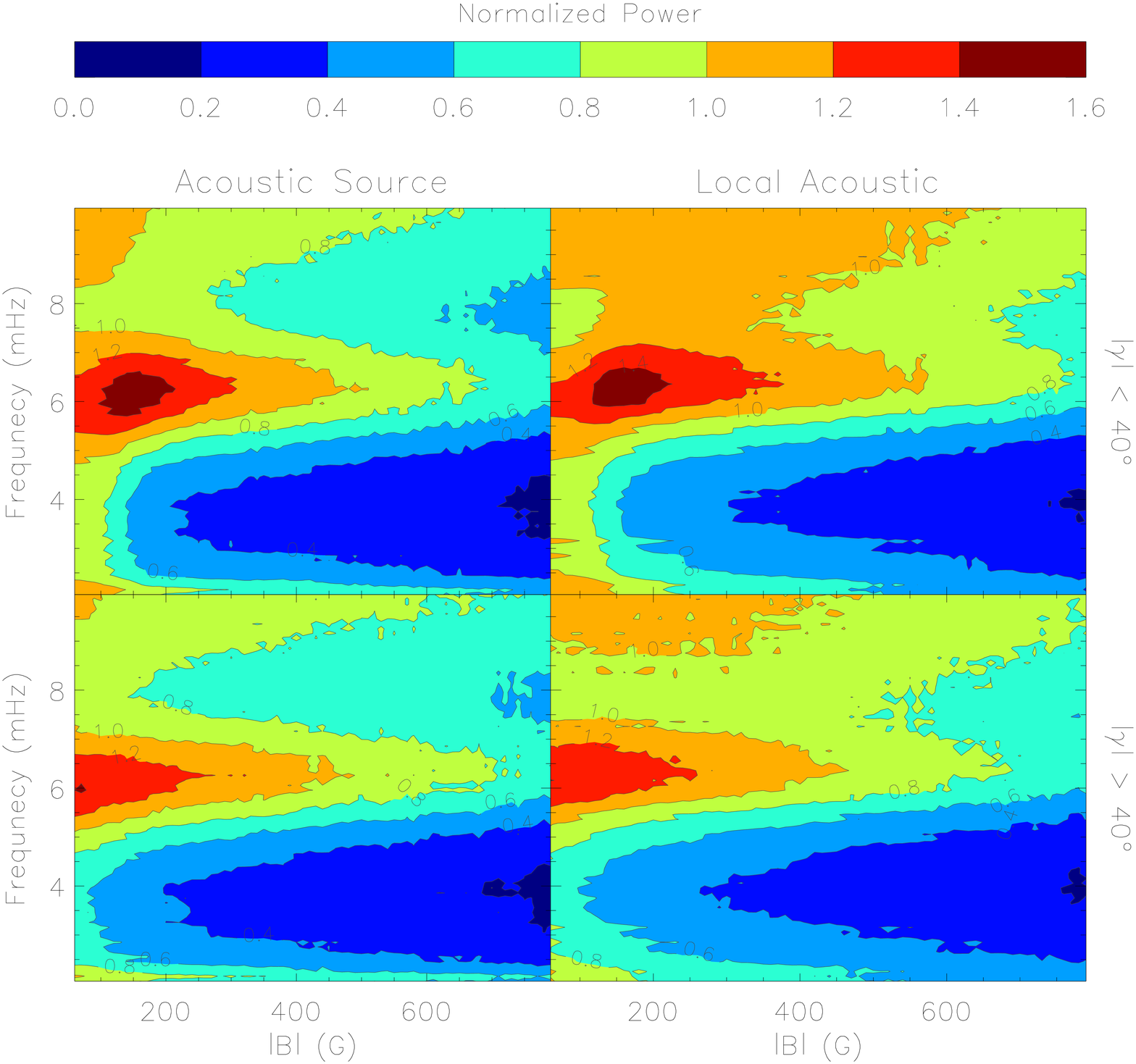}
\caption{The average normalised power as a function of magnetic-field strength for two different ranges of $|\gamma|$ (top: $<40^\circ$, bottom: $>40^\circ$). The left frame shows the acoustic-source power, while the right frame shows the local acoustic oscillation power.}
\label{fig:APEP}
\end{figure}


\section{Results}\label{S-results}
\subsection{Frequency Dependency}
For comparative purposes between the two types of power maps, Figure~\ref{fig:APEP} shows the average power of both the acoustic-source (left) and local oscillations (right) for the entire active region, as a function of field strength, in two inclinations ranges $|\gamma|<40^\circ$ (closer to local horizontal, top) and $|\gamma|>40^\circ$ (closer to local normal, bottom). We divide the full inclination range at $40^\circ$ as the local acoustic power has generally been observed to be enhanced below this limit. In all four frames the average power peaks between 5\ld7\,mHz and at a field strength between 60\ld300\,G.  The frames with a field inclination less than $40^\circ$ have a broader enhanced power distribution with a maximum ($|H_+|^2/|H_{+0}|^2>1.4$) centred at 6\,mHz. Additionally, the acoustic moat is apparent below 4.5\,mHz with a strong power deficit at greater field strengths and inclinations. A comparison between the two types of power maps reveals significant similarities, with subtle differences. While both power maps peak at the same frequency and field strengths, there is more local acoustic power at higher frequencies and field strengths. Additionally, the source power also has a greater power deficit within the frequency range of the acoustic moat.

\subsection{Magnetic Fields in Halos}

\begin{figure}
\centering
\includegraphics[width=\textwidth]{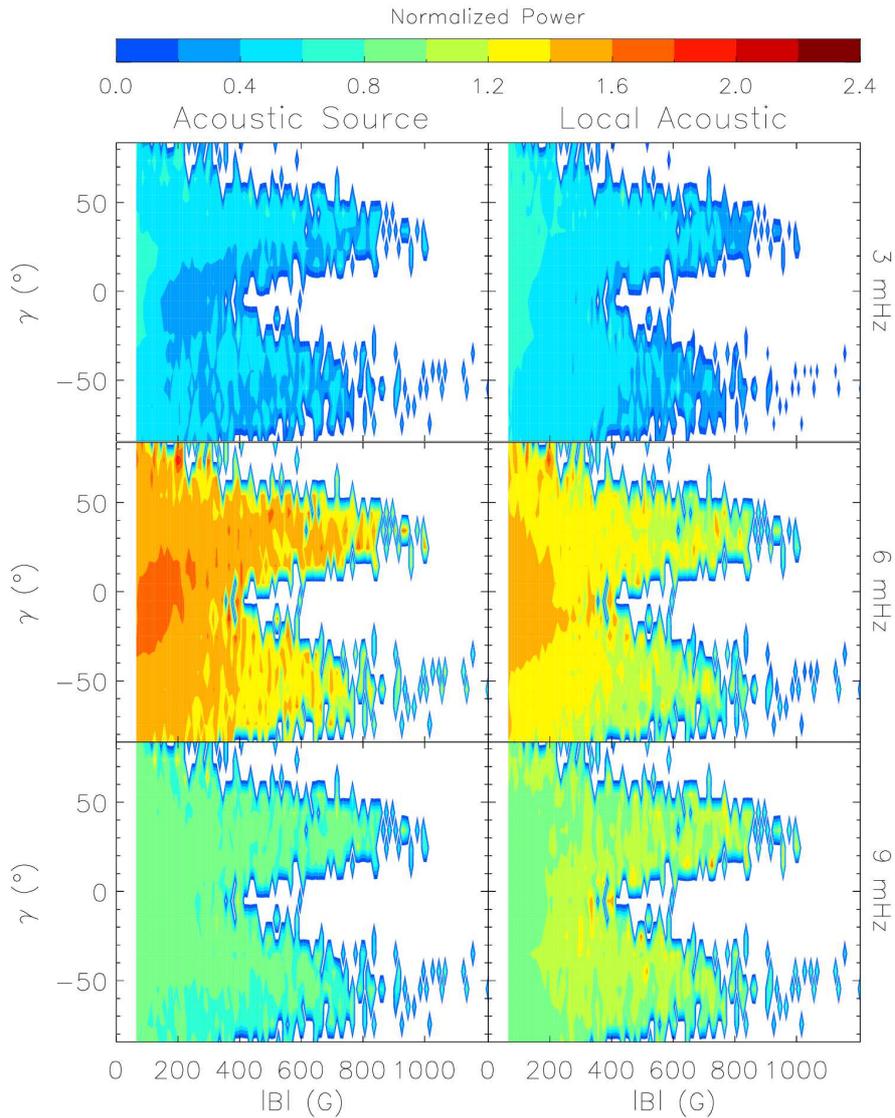}
\caption{The mean normalised power (to the quiet Sun) of the acoustic-source (left) and local acoustic power (right) for 3,6, and 9\,mHz (top to bottom, respectively), within the acoustic-source halo outlined in Figure~\ref{fig:haloandg}.}
\label{fig:inc_b_pow}
\end{figure}

Shifting our focus to the role of magnetic fields on enhanced acoustic-source power, Figure~\ref{fig:inc_b_pow} shows the relationship between the field strength, inclination, and the associated power within the halo. The local/source power has been averaged into bins of 10~G  [$|\bm{B}|$] and $10^\circ$ [$\gamma$]. 
It should also be noted that we are only examining the acoustic-source halo at 6~mHz (yellow contour of Figure~\ref{fig:haloandg}). Hence in all panels the distributions (for specific $\gamma$ and $|\bm{B}|$) are identical, with the corresponding power being the only difference between each frame. Figure~\ref{fig:inc_b_pow} reveals numerous morphological properties of both halos: i) There is a general symmetry of power within the halo around $\gamma=0$. ii) The relationship of power with frequency (shown in Figure~\ref{fig:APEP}) is reflected here, with the greatest enhancement seen at 6~mHz and suppression at 3~mHz. iii) At 3 and 9~mHz the local acoustic power generally dominates over the acoustic-source power (which is dominant at 6~mHz). iv) The greatest acoustic-source power, as well as local acoustic power, occurs within relatively horizontal ($\gamma < 40^\circ$) and fairly weak fields ($<300$G). v) At 9~mHz the acoustic power is enhanced at greater magnetic-field strengths ($>200$G) and more inclined fields ($<60^\circ$), while the acoustic-source power is similar to quiet Sun values.

Again, these results demonstrate the similarities between the local acoustic and acoustic source power's relationship with the magnetic field. However, as stated above there are also notable differences that distinguish the two. Here, the differences in the extent of power enhancement can be seen clearly, as well the halo's behaviour at high frequencies. As a concluding note to this result, there also appears to be an outlier to these observations at 6~mHz, 200G and $\gamma=+80^\circ$ which is greatly enhanced in both power types, relative to its neighbouring results.

\subsection{Statistics of the Zones I, II, and III}

\begin{figure}
\centering
\includegraphics[width=\textwidth]{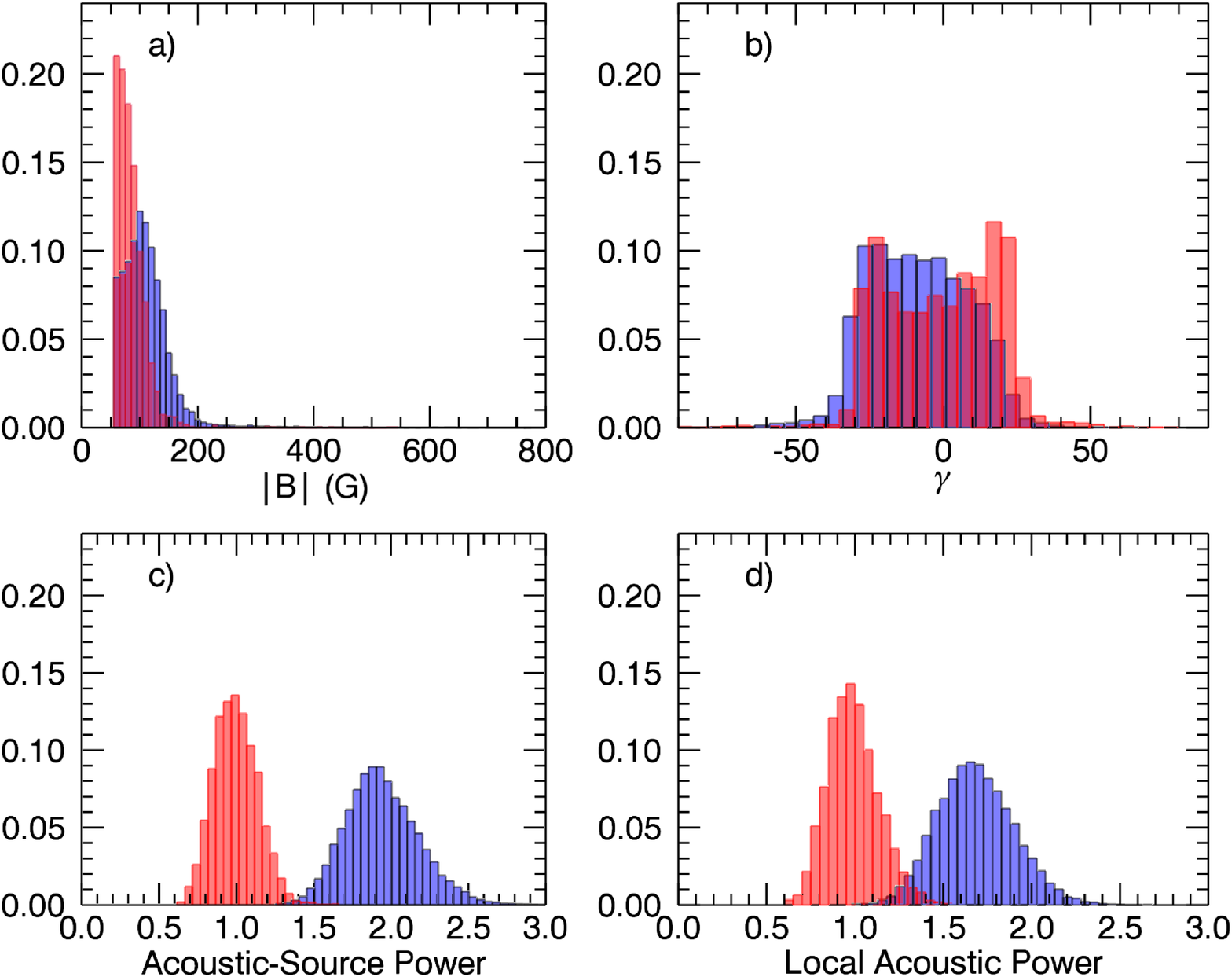}
\caption{Magnetic statistics in the quiet Sun (light red) and acoustic-emission glories (blue):  histogram of magnetic field and inclination (top row),  distribution of acoustic-source density (lower left) and local acoustic-power (lower right). The halo has a mean acoustic source power $90\,\%$ greater then the quiet Sun, while the local acoustic power is  only $70\,\%$ greater.  When compared to Figure~\ref{fig:moathisto} the acoustic glories have a compact distribution over a small magnetic-field strength, while the distribution across field inclination is similar to the quiet Sun.}
\label{fig:gloryhisto}
\end{figure}

Figure~\ref{fig:gloryhisto} shows the distribution of the acoustic glories, as a comparison with a quiet Sun distribution. Panel a shows that the peak magnetic-field strength is 120\,G, while the quiet-Sun magnetic field is at 60\,G. The distribution of magnetic-field strength within the glories is narrow, with the maximum field strength reaching around 200\,G. The distribution of magnetic-field inclination (panel b) within the glories is similar to the quiet-Sun values, which are within $\pm 40^\circ$ of horizontal. Furthermore, panels c and d show that power distribution within the glory is Gaussian, with the mean source power being $190\,\%$ (FWHM\,=\,0.45) more powerful then the quiet Sun, while the local acoustic power is at $170\,\%$ (FWHM\,=\,0.5).

\subsection{The Seismic Plateau}

\begin{figure}
\centering
\includegraphics[width=\textwidth]{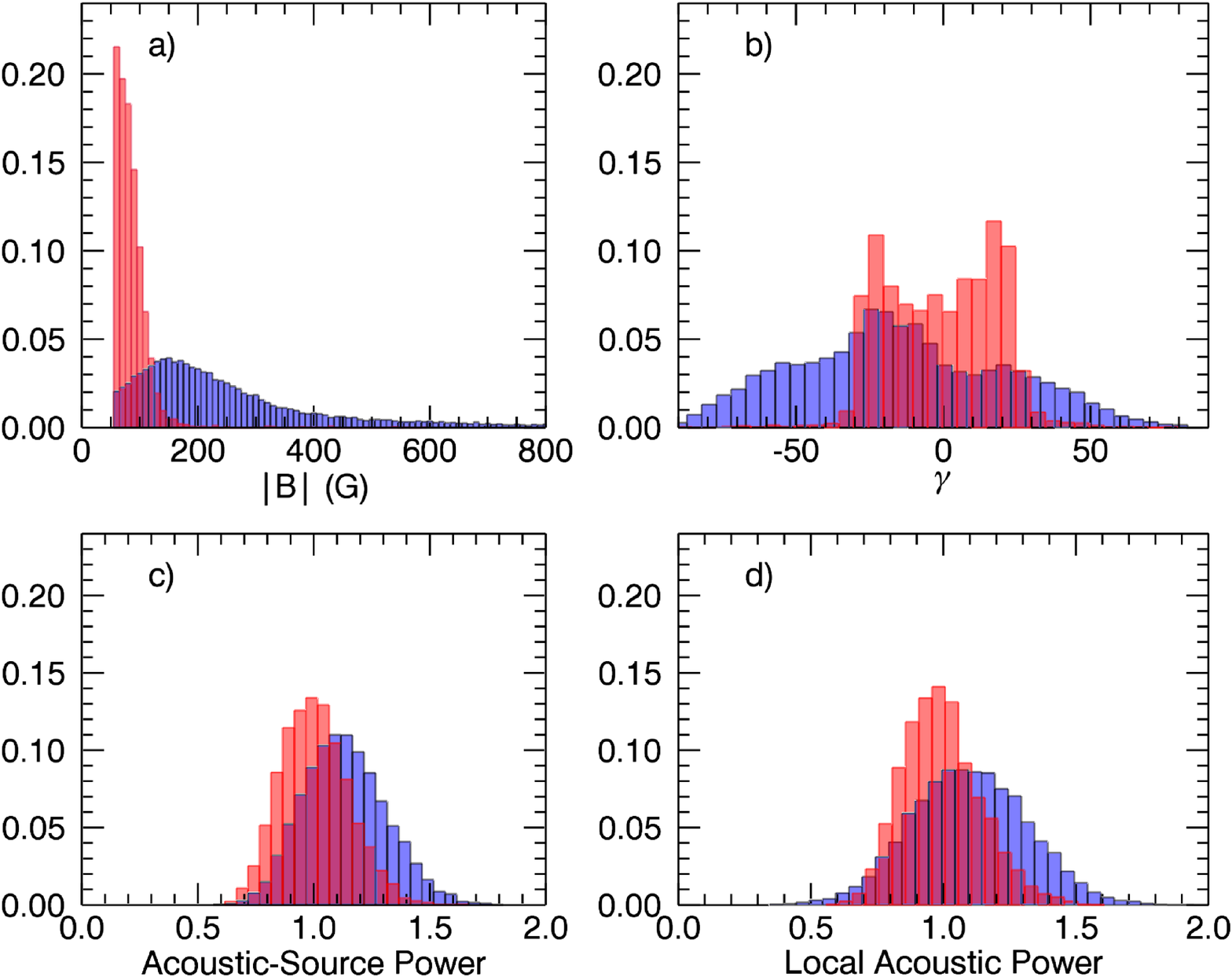}
\caption{Histogram plots of the seismic plateau (blue), compared to the quiet Sun (light red). The histograms are normalized to the number of pixels within both regions. The seismic plateau has a mean local acoustic and acoustic-source power $10\,\%$ larger than the quiet Sun. However, the magnetic-field strength has a broader distribution, as well as the field inclination.}
\label{fig:moathisto}
\end{figure}

The acoustic-source power maps at 6~mHz present a near-quiet-Sun power belt ($\approx 1.1 \times$ quiet-Sun values) situated between the inner boundary of the acoustic-source halo and the external penumbra boundary of the active region (see Figure~\ref{fig:haloandg}). The plateau's width averages between 3 and 12~Mm, with some of the active region's plage contaminating this region. The plateau's width does not depend upon the pupil size in the egression, indicating that the plateau is real and not an artefact of the holography. Figure~\ref{fig:moathisto} shows that while the acoustic-source and local acoustic power have a mean value close to that of the quiet Sun, the magnetic-field morphology differs significantly. The plateau field strength peaks at $150$\,G and has a much broader distribution than the quiet Sun, skewing towards field strengths as great as 400\,G. The field inclination is also broadly distributed across all angles with most field inclinations within $\pm70^\circ$ of horizontal.  The histogram distribution of the source and local acoustic power (Figure~\ref{fig:moathisto} c) and d)) show a slight enhancement (10\,\%) when compared to the quiet Sun. Otherwise, both power values show a similar Gaussian distribution.

\subsection{Spatial Correlation of Maps}
\begin{figure}[!htbp]
\centering
\includegraphics[scale =0.5]{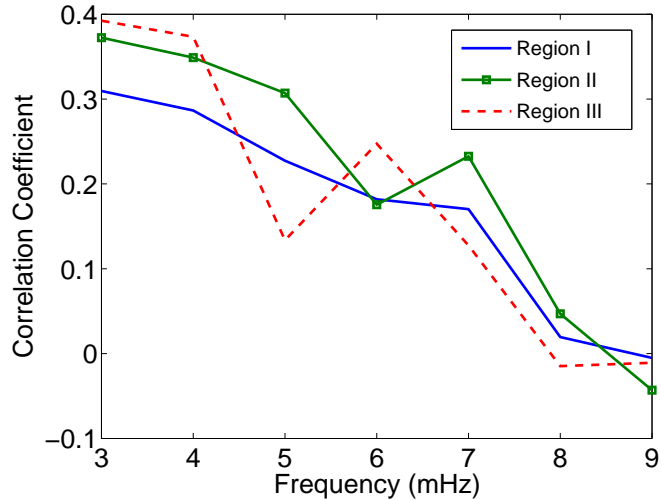}
\caption{The spatial correlation between the highest power regions (at $70\,\%$ of their maximum values) within the glory regions of the local acoustic and acoustic-source, as a function of frequency. In general the correlation between these high powered compact regions is weak, and weakens further with increasing frequency.}
\label{fig:cospatial}
\end{figure}

We conclude this section by examining the spatial correlation between the local acoustic power and the acoustic-source power. An examination of Figure~\ref{fig:powermaps} (and Figure 5 of \citet{donea_hanson_2013}) shows a good general large scale spatial correlation of the highest-power regions within the halo, with these glories occurring within areas situated between high-strength magnetic fields.  However, a close examination of the most powerful regions in each map, with a power greater than $70\,\%$ of their respective peak values, shows a decrease in the correlation with high frequencies (Figure~\ref{fig:cospatial}). The lower limit of $70\,\%$ of peak values corresponds to a power of $124\,\%$ greater than quiet Sun values in the acoustic-source maps and an excess power of $85\,\%$ in the local acoustic maps (see distributions in Figure~\ref{fig:gloryhisto}).

This result is surprising given that the wave field of the egression maps is a subset of the acoustic maps. However, from the analysis thus far this result is consistent with the differing behaviour of the power maps at high frequencies. In these localized high-power regions, where the source power does not correlate with the local acoustic, this may suggest that acoustic waves were generated there (and resurfaced within the pupil), but it is not apparent in the local acoustic maps due to dampening by local acoustic oscillations. Conversely, where the local acoustic power is enhanced, and the source is not, it may indicate the enhancement of local Doppler signatures due to some yet unknown process that generates localized motion. Nevertheless, to further understand this lack of correlation a better understanding of the ``acoustic shower-glass" \citep{lindsey_braun_2005} is needed. 

\section{Discussion and Conclusions}
\label{S-Discussion}

We have used high-resolution SDO/HMI images and vector\ld magnetograms to explore the relationship between the enhanced  acoustic-source and the local acoustic oscillation power observed in the halos of active regions. In this article we have taken significant statistical precautions in examining the weak magnetic fields in the region of the halo. Our main conclusions may be summarized as follows:
\begin{enumerate}[i)]
\item The  complex multi-polar active region AR\,11429 has large patches of enhanced acoustic sources localized mainly in magnetically ``quiet'' areas surrounding the sunspot penumbrae.
\item We have been able to determine accurate statistical relationships between the enhanced acoustic sources and the local magnetic-field distribution;
\item The enhanced acoustic sources occur at 5\ld7\,mHz within regions of intermediate field strengths (60\ld200~G), but with inclinations no different than that found in the quiet Sun;
\item A ``plateau" of intermediate acoustic-source power is recognised as a transition region between the active region and halo, in which the local acoustic and acoustic source power is suppressed;
\item Our results confirm those of previous studies, showing that the acoustic wave field is enhanced within horizontal fields ($\pm 40^\circ$);
\item The morphology of the acoustic-source halo and local acoustic halo are similar in many ways, but differ at high frequencies and in the degree of power enhancement;
\item Enhanced acoustic glories occur in regions of intermediate field strength;
\item There is not a strong spatial correlation  between the local acoustic power and the enhanced acoustic sources in the halos;
\end{enumerate}

Our results complement those of previous studies \citep{schunker_braun_2011,rajaguru_etal_2012}, which agree that a relationship exists between the enhanced local acoustic power and the magnetic field. However, unlike those studies, we are able here to differentiate acoustic sources from other acoustic oscillations, through the use of holography. We have thoroughly examined the magnetic morphology of three distinct regions of acoustic emission: the halo, glories and what we refer to as the seismic plateau. In examining these regions we have found that the enhancement of acoustic-sources is also greatly influenced by the magnetic field, with enhanced emission occurring within regions of quiet-Sun-like field inclinations. However, regions interwoven with magnetic plage (such as the plateau) hinder acoustic emission, producing near-quiet-Sun power belts situated between the enhanced halo and the strong magnetic fields of the active region.

With regards to the relationship between the local acoustic power and the acoustic-source power, we find strong similarity. Both halos show the greatest enhancement in the same frequency range and magnetic-field morphology. The similarity in these results is to be expected considering that the sources computed by holography are a subset of the wave field that produces the local acoustic wave field.  However, examination of spatial correlations show that localized acoustic noises may hinder high-powered local acoustic kernels, that are otherwise seen in acoustic-source maps. Furthermore, it is possible that local emitters submerged just below the photosphere emit different amounts of energy in an upwards direction, when compared to the downwards direction. For example, if the energy upward (contributing to the local oscillations) was different from that emitted downward (contributing to the signals sampled in the pupil) then one would expect to see the power differences reported in this study. We understand this would not be the case if the emitters were simple dipoles or quadrupoles, but it is possible to contrive coherently related combinations of dipoles and quadrupole emitters that support this. 

\citet{rajaguru_etal_2012} recently calculated the local acoustic power halo at differing chromospheric heights, concluding that the radius of the halo increases with height. Currently, we are restricted by the methodology of holography to calculating acoustic-sources below the photospheric surface. Consequently, we cannot make direct comparisons with the halos at chromospheric heights. However, the depth-diagnostic abilities of holography may warrant a comparison with numerical models that can calculate local acoustic power maps below the surface.

While we have shed further light on the acoustic sources surrounding a complex active region, the question of what mechanism is responsible is still one for further examination. Our results of the magnetic-field morphology and power enhancement are consistent with processes that require an interaction with the horizontal field of a canopy structure \citep{muglach_etal_2005}, including mode-conversion \citep{khomenko_collados_2009} and the trapping of waves under the canopy \citep{kuridze_etal_2008}.

With regards to the trapping of waves, \citet{kuridze_etal_2008} suggested that a small-scale canopy above a solar granules (on scales of $\approx0.5$\,Mm) will create a field-free cavity below, in which waves are trapped and enhance the local Doppler signature. While the model was very simple (cold-plasma, unstratified \textit{etc.}), one would expect to see evidence of this process as acoustic sources restricted to within granular regions. The diffraction-limited (2.5\,Mm) maps of holography are unable to resolve these small emitters. However, to generate the observed acoustic halo and glories, these small cavities would have to be located closely altogether, be broadly distributed, and have lifetimes on a par with the observation time frame. With regards to the mode conversion process, fast waves with frequencies above the acoustic cut-off travel to chromospheric heights to be refracted and converted into slow-waves. This process is sensitive to field inclinations \citep{cally_2006}, and should generate additional acoustic sources in horizontal-field regions, where a majority of the fast wave is returned to the surface. The change in the morphology of local acoustic halos with chromospheric height \citep{rajaguru_etal_2012} complements this theory, and our analysis of the acoustic-source halos confirms that enhancement is significant within regions of near-horizontal inclinations, as suggested by mode conversion. We note that recent forward modelling work by \citet{rijs_etal_2015} presents results suggesting fast-wave refraction can produce the observed acoustic-sources. 

\citet{hindman_brown_1998} suggested that kink motion of magnetic-flux tubes could deposit additional energy into the photosphere. This addition of energy into regions already enhanced by canopy-related mechanisms could give rise to glories. However, this concept presents two problems that will need addressing: Firstly, it is unlikely that the kink mode is responsible for any observed additional vertical oscillations, as it is transversal in nature and any motion of the plasma will not be seen in Doppler images near the disk centre. We suggest that if tube motion is responsible, then it is most likely  sausage motion, in which the compression of plasma would add to the vertical-velocity profile. The second issue is a question of flux-tube lifetimes. Anchored flux tubes (in the surrounding photosphere) are generally located at granular boundaries and any motion of these tubes would have the same lifetime as the solar granules (on order of minutes), adding little to the power maps over 24 hours. We suggest the presence of acoustic glories is probably due to some small scale dynamical process, that has yet to be understood in terms of sustained enhanced acoustic emission.

While we have discussed the agreement of our results with some proposed mechanisms, there need to be further observational studies, into the nature of the magnetic fields, the plasma-$\beta$ heights, magnetic canopies and their relationship with the oscillation and source power halo. Furthermore, theoretical and numerical models need to be developed, improved, and tested against observable parameters to better understand the mechanism responsible for the enhanced acoustic sources surrounding complex active regions.


\section*{Conflict of Interest}
The authors declare that they have no conflict of interest.

 \begin{acks}
The data used here are courtesy of NASA/SDO and the HMI science team.
Data analysis was performed on the Monash University SunGrid. Research performed by Chris S. Hanson at NWRA was funded by the Monash University Institute of Graduate Research.
 \end{acks}


 \bibliographystyle{spr-mp-sola}
\bibliography{../../../BibTex/References}

\end{article} 
\end{document}